\documentclass[10pt,twocolumn,letterpaper]{article}

\usepackage[pagenumbers]{cvpr} 

\usepackage[dvipsnames]{xcolor}

\usepackage{multirow}

\definecolor{cvprblue}{rgb}{0.21,0.49,0.74}
\usepackage[pagebackref,breaklinks,colorlinks,citecolor=cvprblue]{hyperref}

\def\paperID{11168} 
\def\confName{CVPR}
\def\confYear{2024}

\title{Test-Time Backdoor Defense via Detecting and Repairing}

\author{%
  Jiyang Guan$^{1,2}$, ~Jian Liang$^{1,2}$,~Ran He$^{1,2}$\\
  $^{1}$MAIS $\& $CRIPAC, Institute of Automation, Chinese Academy of Sciences, China \\
  $^{2}$School of Artificial Intelligence, University of Chinese Academy of Sciences, China\\
  {\tt\small guanjiyang2020@ia.ac.cn,
\tt\small 	liangjian92@gmail.com, \tt\small rhe@nlpr.ia.ac.cn}
}

\begin{document}
\maketitle
\begin{abstract}
Deep neural networks have played a crucial part in many critical domains, such as autonomous driving, face recognition, and medical diagnosis. 
However, deep neural networks are facing security threats from backdoor attacks and can be manipulated into attacker-decided behaviors by the backdoor attacker.
To defend the backdoor, prior research has focused on using clean data to remove backdoor attacks before model deployment. 
In this paper, we investigate the possibility of defending against backdoor attacks at test time by utilizing partially poisoned data to remove the backdoor from the model.
To address the problem, a two-stage method Test-Time Backdoor Defense (TTBD) is proposed.
In the first stage, we propose a backdoor sample detection method DDP to identify poisoned samples from a batch of mixed, partially poisoned samples.
Once the poisoned samples are detected, we employ Shapley estimation to calculate the contribution of each neuron's significance in the network, locate the poisoned neurons, and prune them to remove backdoor in the models.
Our experiments demonstrate that TTBD removes the backdoor successfully with only a batch of partially poisoned data across different model architectures and datasets against different types of backdoor attacks.
Code is attached in the supplementary material.
\end{abstract}    
\section{Introduction}
\label{sec:intro}
Over the past years, deep neural networks have played a crucial part in many critical domains, such as autonomous driving \cite{tian2018deeptest}, face recognition \cite{taigman2014deepface}, and medical diagnosis \cite{kermany2018identifying}.
Despite their widespread use, deep neural networks lack transparency, making them vulnerable to different attacks. 
This vulnerability leads to serious mistakes in security-related areas, causing significant threats and concerns. 
Recent studies have proven that backdoor attacks \cite{liu2023detecting} pose a severe security threat to deep neural networks.
Backdoor attacks \cite{gu2017badnets}, taking advantage of the overfitting of the deep neural networks, inject the backdoor poisoned data with the small invisible triggers into the model's training dataset, and cause the model trained on that to behave normally on clean samples but predict the wrong, attacker-decided labels on the backdoor poisoned samples.
A wide variety of backdoor attacks \cite{gu2017badnets,chen2017targeted,barni2019new,zeng2021rethinking,nguyen2021wanet} have been proposed, and recent works such as WaNet \cite{nguyen2021wanet} and LF \cite{zeng2021rethinking} add invisible backdoor triggers onto clean samples, leading to more serious security threats.

To reduce the threats of backdoor attacks, numerous backdoor defense methods have been proposed, which can generally be classified into two categories: the training-stage backdoor defense and the post-training backdoor defense (model repairing backdoor defense).
Because training deep neural networks is an expensive process that requires significant data collection and computational resources, training on the cloud and directly using third-party well-trained models have become increasingly popular, and thus, in this paper, we mainly focus on the model repairing backdoor defense.
The model repairing backdoor defense usually involves retraining \cite{wang2019neural} or pruning \cite{liu2018fine,wu2021adversarial} to remove backdoor based on the clean data, for example, by using $5\% $ of clean samples from the training dataset.
Previous works have mainly focused on repairing poisoned models before model deployment, while in this work, we investigate whether the backdoor defense can be performed during test time by the defender.

At test time, the defenders have access to the mixture of both the clean samples and the backdoor poisoned samples, or in other words, the partially poisoned data.
While the poisoned samples in the mixed, partially poisoned data can help remove backdoor, detecting them accurately poses a significant challenge. 
Moreover, since poisoned sample detection can not be entirely accurate, there may be some clean samples mixed with the detected poisoned samples, which presents another challenge for backdoor removal.
As for the backdoor poisoned sample detection, TeCo \cite{liu2023detecting} is a state-of-the-art backdoor sample detection method that leverages sample corruption robustness consistency to distinguish between poisoned and clean samples. 
However, TeCo is sensitive to different model architectures and can not successfully distinguish poisoned data from clean data on some model architectures such as VGG \cite{simonyan2014very}, causing problems to backdoor removal. 
To overcome this problem, we propose a novel backdoor detection method called Detection During Pruning (DDP), which can accurately detect poisoned samples across different model architectures. 
After poisoned data detection, we propose a Shapley-based backdoor cleanse method, which can tolerate imprecise backdoor detection. 
Our experiments demonstrate that our two-stage test-time backdoor defense framework Test-Time Backdoor Defense (TTBD) can remove backdoor using only a small batch of partially poisoned data (100 images) and remove backdoor in models with only a small decrease in accuracy across 3 common model architectures and 3 common datasets facing 7 different types of backdoor attacks.

Our contributions are summarized as follows:
\begin{itemize}

\item We introduce a realistic scenario of removing the backdoor from models at test time with partially poisoned data, offering a new perspective on backdoor defense.

\item We propose a novel two-stage test-time backdoor defense framework, TTBD,  which detects the poisoned samples at test time and then uses Shapley estimation to guide poisoned neuron pruning.




\item The experiments demonstrate that TTBD removes backdoor successfully with only a batch of partially poisoned data across different model architectures and datasets facing various attacks.
\end{itemize}

\begin{figure*}
\centering
\includegraphics[width=0.85\textwidth]{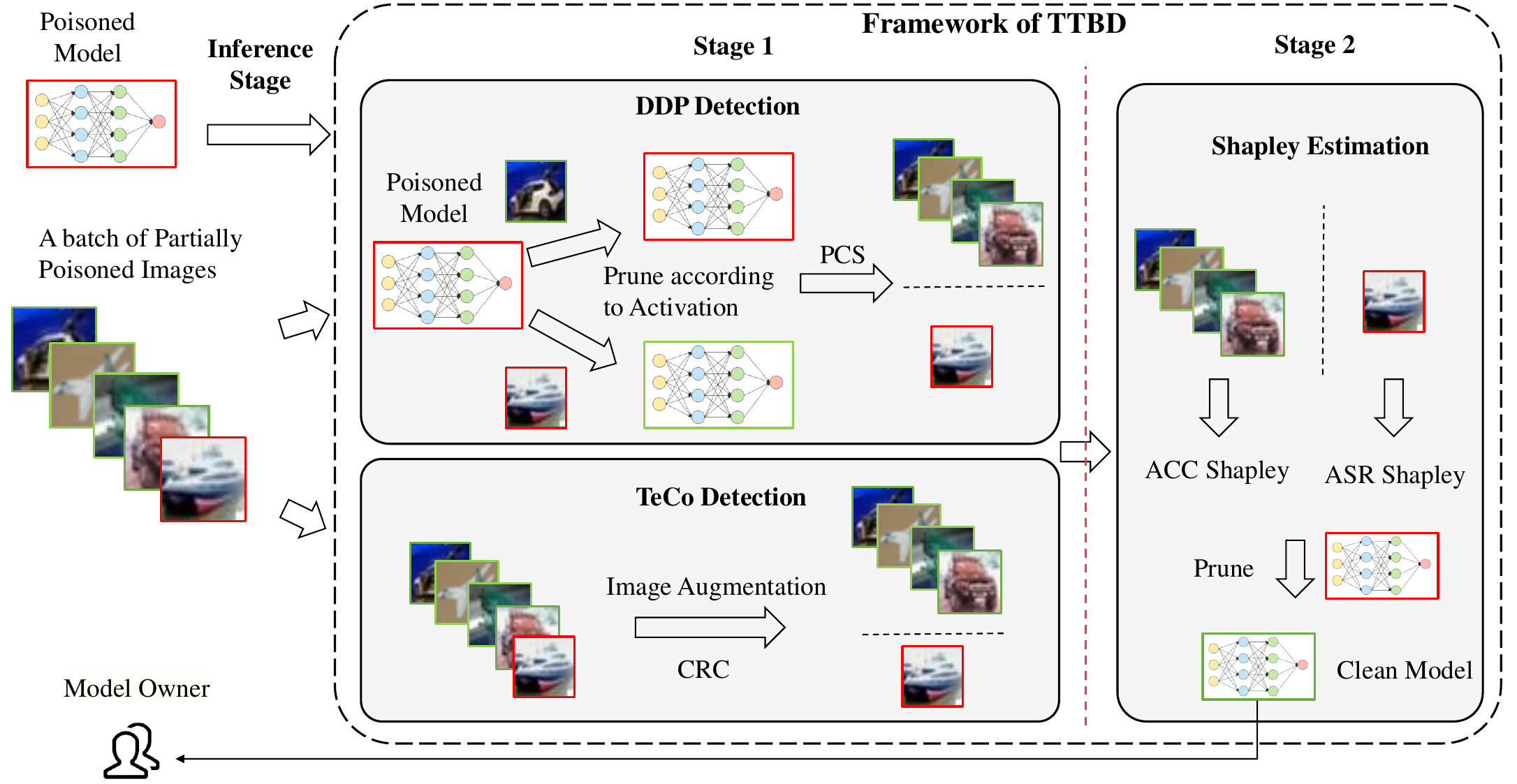}
\centering
\caption{Framework of Test-Time Backdoor Defense (TTBD). Samples and models in the red broader represent poisoned samples and models, and samples and models in the green broader represent clean samples and models.}
\vspace{-4mm}
\label{fig:TTBD}
\end{figure*}
\section{Related Work}
\subsection{Backdoor Attack}
Backdoor attacks usually inject poisoned label-flipped samples with the attacker-decided backdoor triggers into the model training dataset and cause the poisoned model trained on that to behave normally on clean samples but be manipulated on the poisoned samples.
The first and most famous backdoor attack is BadNets \cite{gu2017badnets}, which injects a small square at the corner of the image as the backdoor trigger, leading to the model's misclassification on the triggered samples.
To make the backdoor trigger more invisible, the following works leverage strategies such as blending \cite{chen2017targeted}, natural reflection \cite{liu2020reflection}, low frequency \cite{zeng2021rethinking} and encoder-decoder framework \cite{li2021invisible} to design the invisible backdoor triggers. 
Furthermore, to make backdoor attacks more flexible and convert, the multi-target and multi-triggers attacks have been proposed \cite{xue2020one,zheng2022pre}.

\subsection{Training-Stage Backdoor Defense}
Under this setting, the defenders have access to the training dataset and are able to control the model training process, so that they can detect and filter the poisoned data or add restrictions to suppress the overfitting of backdoor during the training process \cite{zheng2022pre}.
The backdoor samples, as outliers of the training dataset, have different representation statics in the feature space and different sensitivity to image transformation, and thus the defenders make use of these differences to filter out the poisoned samples in the training dataset \cite{doan2020februus,huang2022backdoor,huang2022distilling}.
Other methods leverage restriction to model training to weaken the influence of backdoor samples during the training process \cite{li2021anti,rosenfeld2020certified,steinhardt2017certified}.

\subsection{Post-Training Backdoor Defense}
With the rise of machine learning as a service (MLass) in recent years, users can leverage third-party models directly without training models by themselves.
However, these models may be poisoned, which poses a significant threat to the users.
Under such circumstances, defenders need to remove backdoors from the models without having access to the training dataset or training process.
Most post-training backdoor defense methods leverage clean labeled data to prune backdoor neurons \cite{liu2018fine,guan2022few}, reverse the backdoor trigger and unlearn it \cite{wang2019neural,zhu2020gangsweep,guo2020towards,chen2019deepinspect}, adversarially activate backdoor neurons \cite{wu2021adversarial}, and distillate neural attention \cite{lineural} to remove backdoor.
All these methods defend against the backdoor before the model deployment and neglect the backdoor defense during the test-time stage.
During test time, the defender can only get access to the partially poisoned data and how to use it to remove the backdoor in the deployed models is still a problem.

There are a few sample-oriented backdoor defense methods during test time \cite{li2022test,sun2023mask}.  
However, both of them only focus on eliminating triggers from samples and do not repair the backdoor models, which causes the model to process detection or elimination on each sample and affects the model's inference speed a lot.
Additionally, sample-level trigger elimination requires a pre-trained autoencoder, which restricts its applicability and scope of use. 
Since backdoor behavior is typically due to the overfitting of the backdoor trigger and the specific target label in models, repairing at the model level is generally more effective. 
Our test-time backdoor defense method removes the backdoor by repairing the model using only one batch of test-time partially poisoned samples.

\subsection{Test-Time Adaptation}
Test-Time Adaptation (TTA) \cite{liang2023comprehensive} holds significant importance within the area of Domain Adaptation (DA).
In some situations where source training data is not available, Test-Time Adaptation leverages test-time data in the target domain to adapt the models without access to training data, garnering considerable attention.
Generally, existing TTA methods make use of strategies such as mutual information maximization and pseudo-label training \cite{liang2020we}, and entropy minimization \cite{wang2020tent} to adapt to the target domain.
In the realm of defense against attacks, several test-time adversarial defense methods have been proposed. 
The majority of these methods rely on detection and sample purification techniques \cite{nayak2022dad,croce2022evaluating,nayak2023dad++,perez2021enhancing} to eliminate adversarial noise from samples during testing. 
Notably, these methods achieve noise removal without requiring clean data for adversarial training, significantly reducing the demands on defenders.
Similar to TTA and test-time adversarial defense, removing the backdoor from the pre-trained models during test time is also important.
Especially, in some cases, only the potentially partially poisoned test data is available.

\section{Proposed Method}
\subsection{Problem Definition}
We consider a realistic scenario in which the defender seeks to remove backdoor attacks in models at test time. 
In this scenario, the attackers attempt to use the poisoned data to attack the defenders' backdoor models, and intuitively, the defenders are able to make use of those poisoned data to remove the backdoor in their models. 
However, the attackers typically do not use the fully poisoned data to avoid suspicion of the defenders.
Thus in most cases, the defenders are provided partially poisoned data which mixes the clean samples and the poisoned samples. 
Then how the defenders use the partially poisoned data to remove the backdoor is a challenging problem.

\subsection{Poisoned Sample Detection}
During test time, the defenders only have access to partially poisoned samples and we hope the defenders remove the backdoor from the models with only a batch of partially poisoned images. 
To solve the test-time backdoor removal problem, we propose a two-stage backdoor defense method. 
Intuitively, the first step of backdoor removal is to use poisoned sample detection to distinguish the poisoned samples from clean samples.
And then, the defenders leverage these detected poisoned samples to remove the backdoor from the backdoor models.
An overview of our framework is shown in Figure \ref{fig:TTBD}.
In this subsection, we focus on the first stage of TTBD.
To be specific, we review a SOTA detection method TeCo, and propose our backdoor sample detection method DDP.

\paragraph{$\triangleright$ Detection with TeCo.}
TeCo \cite{liu2023detecting} is a state-of-the-art backdoor detection method, which utilizes samples' corruption robustness consistency to distinguish between the poisoned samples and the clean samples. To be specific, TeCo conducts Corruption Robust Consistency (CRC) as the detection indicator, expressed as:
\begin{equation}
\begin{aligned}
   CRC = deviation(Severity),
   \\ Severity=[sev_{1},\cdots sev_{i}, \cdots sev_{n}]
\end{aligned}
\end{equation}
where $sev_{i}$ represents the recorded severity for the i-th type of corruption. 
However, TeCo has some limitations, such as being sensitive to different model architectures. 
It fails when faced with certain model architectures, such as VGG. 
To deal with the problem, we propose a new poisoned sample detection method called DDP.

\paragraph{$\triangleright$ Detection with DDP.}
The previous backdoor detection method TeCo has used the difference in sensitivity between poisoned and clean samples to image transformations as a means of distinguishing between them.
According to previous works \cite{liu2018fine}, another difference between poisoned and clean samples is that poisoned samples activate backdoor neurons, while clean samples do not.
Therefore, the defenders can make use of the difference in backdoor activation to distinguish the poisoned and clean samples.
Pruning neurons with the highest activation value among the poisoned samples will help remove backdoor in models and when the defenders are given a batch of partially poisoned samples, they can use the activation of each sample in the batch to guide the pruning process and remove backdoor attacks in the model.
For poisoned samples, pruning neurons with the highest activation value results in a sharp decrease in the model's ASR, leading to changes in predictions for both normal and poisoned samples. However, for clean samples, pruning neurons with the highest activation value only causes changes in the normal samples. Therefore, we use the prediction change score (PCS) as an indicator of poisoned sample detection. Poisoned samples typically exhibit a higher PCS than clean samples, and PCS is expressed as:
\begin{equation}
   PCS = \sum\limits_{i} (\hat{prediction_{i}} \neq prediction_{i})
\end{equation}
where $prediction_{i}$ represents the original model's prediction on the i-th sample and $\hat{prediction_{i}}$ represents the pruned model's prediction on the i-th sample.
To prevent the accuracy of the model from influencing the PCS, pruning is early stopped when the accuracy of the model drops below a certain threshold.
Pruning based on samples' activation value is an intuitive way to remove backdoor in models and detect poisoned samples.
However, in our setting, only one sample is used at a time for activation estimation and poisoned neuron location, which reduces the performance of backdoor removal using the poisoned samples' activation. 
To address this limitation, we calculate the Shapley value of the neurons using the whole batch of partially poisoned samples as the indicator of neurons which are important to both clean and backdoor samples and prune the neurons with the top-k activation values and the bottom-l Shapley values. 
This approach provides a more accurate measure of the importance of each neuron in the neural network and can effectively locate the poisoned neurons. 
The calculation of the Shapley value and its effectiveness will be discussed in the following section.

\paragraph{$\triangleright$ How to detect under extremely low poising rates?}
\label{TTBD-SPARSE}
In some cases, the attackers may feed the backdoor samples sparsely (some of the batches are totally clean) during the inference time or attack at a very low poisoning rate to avoid the defenders' detection.
In these cases, defenders have the capacity to aggregate images from multiple batches and subsequently remove the backdoor.
When confronted with such minimal instances of poisoning, defenders leverage a combination of techniques including TeCo and DDP as a dual detection approach to enhance the accuracy of detection.
Then the defenders leverage the TTBD framework to remove the backdoor from the models. 
TTBD-SPARSE in Table \ref{tab:performance_sparse} demonstrates that the defenders remove backdoor successfully by aggregating images from 20 batches of images when the attackers attack with a very low poisoning rate $1\%$  and sparsely (some of the batches are totally clean).

\begin{table*}[h]\footnotesize
  \centering  
 
    \begin{tabular}{l c c c c c c c c c c c c}  
    \hline  
    {Attack}&  
    \multicolumn{2}{c}{Before}&\multicolumn{2}{c}{FP \cite{liu2018fine}} &\multicolumn{2}{c}{ANP \cite{wu2021adversarial} } &\multicolumn{2}{c}{DBD \cite{huang2022backdoor}} &\multicolumn{2}{c}{TTBD-TeCo}  &\multicolumn{2}{c}{TTBD-DDP}

   
    \cr\cline{2-13}  
    ($\%$)&ACC&ASR&ACC$\uparrow$&ASR$\downarrow$&ACC$\uparrow$&ASR$\downarrow$&ACC$\uparrow$&ASR$\downarrow$&ACC$\uparrow$&ASR$\downarrow$&ACC$\uparrow$&ASR$\downarrow$\cr 
    \hline  
    BadNet \cite{gu2017badnets} & $91.23$  & $90.22$ & $91.58$ & $51.52$ & $86.09$ & $1.79$ &$78.09$ & $2.99$&$88.57$ & $1.17$ &$88.50$ & $2.51$ \cr
    Blended \cite{chen2017targeted}   & $93.76$  & $94.88$ & $93.48$ & $94.14$ & $86.92$ & $37.01$ &$70.18$ & $8.04$& $86.00$ & $3.00$ &$88.53$ & $2.24$ \cr
    SIG \cite{barni2019new}  & $91.45$  & $91.47$ & $92.17$ & $94.57$ & $86.26$ & $25.04$&$75.01$& $67.82$&  $88.42$ & $2.17$ &$89.59$ & $2.77$ \cr
    LF \cite{zeng2021rethinking}  & $93.76$  & $86.74$ & $93.30$ & $87.37$ & $84.76$ & $17.78$& $79.13$ & $7.47$ & $90.28$ & $2.05$ &$90.47$ & $2.72$ \cr
    WaNet \cite{nguyen2021wanet}  & $91.48$  & $89.91$ & $91.55$ & $0.20$ & $88.36$ & $0.66$ & $80.90$ & $6.61$ & $91.58$ & $0.49$ &$91.07$ & $0.78$ \cr 
   IAB \cite{nguyen2020input} & $91.09$  & $90.61$ & $92.15$ & $82.06$ & $90.68$ & $4.09$  & $69.66$ & $14.36$ & $88.69$ & $1.77$ &$88.47$ & $1.97$ \cr
    SSBA \cite{li2021invisible}  & $93.43$  & $73.44$ & $93.34$ & $52.76$ & $91.68$ & $2.19$  & $78.52$ & $1.13$ & $91.94$ & $0.17$ &$91.09$ & $3.28$ \cr 
    \hline
     Average & $92.31$  & $88.18$ & $92.51$ & $66.09$ & $87.82$ & $12.65$  & $75.93$ & $15.49$ & $89.35$ & $1.55$ &$89.67$ & $2.32$ \cr\hline
    
    \end{tabular}  
    
    \caption{Defense methods against common attacks on PreAct-ResNet18 using CIFAR10.}
    \label{tab:performance_CIFAR10-PreAct-ResNet}  
\end{table*}

\begin{table*}[h]\footnotesize
  \centering  

    \begin{tabular}{l c c c c c c c c c c c c}  
    \hline  
    {Attack}&  
    \multicolumn{2}{c}{Before}&\multicolumn{2}{c}{FP \cite{liu2018fine}} &\multicolumn{2}{c}{ANP \cite{wu2021adversarial} } &\multicolumn{2}{c}{DBD \cite{huang2022backdoor}} &\multicolumn{2}{c}{TTBD-TeCo}  &\multicolumn{2}{c}{TTBD-DDP}
   
    \cr\cline{2-13}  
    ($\%$)&ACC&ASR&ACC$\uparrow$&ASR$\downarrow$&ACC$\uparrow$&ASR$\downarrow$&ACC$\uparrow$&ASR$\downarrow$&ACC$\uparrow$&ASR$\downarrow$&ACC$\uparrow$&ASR$\downarrow$\cr 
    \hline  
    BadNet \cite{gu2017badnets}& $90.53$  & $84.45$ & $90.35$ & $79.58$ & $-$ & $-$ & $57.54$ & $2.48$& $88.50$ & $84.97$ &$87.11$ & $1.63$ \cr
    Blended  \cite{chen2017targeted} & $90.81$  & $88.31$ & $90.62$ & $87.90$ & $-$ & $-$& $55.24$ & $6.21$ & $86.78$ & $89.73$ &$86.97$ & $4.81$ \cr
    SIG  \cite{barni2019new}& $90.75$  & $82.93$ & $90.60$ & $83.46$ & $-$ & $-$ & $55.92$ & $0.00$& $88.38$ & $83.95$ &$89.48$ & $2.64$ \cr
    LF \cite{zeng2021rethinking} & $88.88$  & $93.43$ & $88.79$ & $92.39$ & $-$ & $-$& $56.83$ & $10.21$ & $86.61$ & $88.18$ &$87.04$ & $1.68$ \cr
    WaNet \cite{nguyen2021wanet} & $89.29$  & $85.63$ & $89.67$ & $72.07$ & $-$ & $-$& $55.80$ & $15.22$ & $88.24$ & $1.38$ &$88.58$ & $1.82$ \cr 
   IAB \cite{nguyen2020input} & $89.24$  & $71.79$ & $89.73$ & $77.43$ & $-$ & $-$  & $57.21$ & $13.10$ & $86.15$ & $20.30$ &$84.77$ & $3.92$ \cr
    SSBA \cite{li2021invisible} & $89.66$  & $91.28$ & $89.59$ & $88.97$ & $-$ & $-$  & $58.45$ & $12.20$ & $87.51$ & $90.63$ &$89.02$ & $1.98$ \cr 
    \hline
     Average & $89.88$  & $85.40$ & $89.91$ & $83.11$ & $-$ & $-$  & $56.71$ & $8.49$ & $87.45$ & $65.59$ &$87.57$ & $2.64$ \cr\hline
    
    \end{tabular}  
   
    \caption{Defense methods against common attacks on VGG19 using CIFAR10.}
     \vspace{-3mm}
    \label{tab:performance_CIFAR10-VGG}  
\end{table*}

\begin{table*}[h]\footnotesize
  \centering  

    \begin{tabular}{l c c c c c c c c c c c c}  
    \hline  
    {Attack}&  
    \multicolumn{2}{c}{Before}&\multicolumn{2}{c}{FP \cite{liu2018fine}} &\multicolumn{2}{c}{ANP \cite{wu2021adversarial}} &\multicolumn{2}{c}{DBD \cite{huang2022backdoor}} &\multicolumn{2}{c}{TTBD-TeCo}  &\multicolumn{2}{c}{TTBD-DDP}
   
    \cr\cline{2-13}  
    ($\%$)&ACC&ASR&ACC$\uparrow$&ASR$\downarrow$&ACC$\uparrow$&ASR$\downarrow$&ACC$\uparrow$&ASR$\downarrow$&ACC$\uparrow$&ASR$\downarrow$&ACC$\uparrow$&ASR$\downarrow$\cr 
    \hline  
    
    BadNet \cite{gu2017badnets}& $55.17$  & $99.91$ & $50.13$ & $99.86$ & $49.83$ & $0.07$ & $44.12$ & $98.89$& $49.42$ & $2.83$ &$53.24$ & $0.25$ \cr
    Blended  \cite{chen2017targeted} & $55.03$  & $99.78$ & $49.75$ & $99.39$ & $50.38$ & $96.02$& $44.51$ & $100.00$ & $47.86$ & $3.75$ &$48.92$ & $7.89$ \cr
    LF  \cite{zeng2021rethinking}& $54.93$  & $98.91$ & $50.37$ & $98.26$ & $49.48$ & $94.44$& $43.80$ & $98.67$ & $51.43$ & $1.81$ &$50.51$ & $0.77$ \cr
    WaNet \cite{nguyen2021wanet}& $46.98$  & $98.00$ & $48.35$ & $73.03$ & $44.98$ & $1.36$& $44.00$ & $99.71$ & $46.92$ & $0.48$ &$47.00$ & $0.60$ \cr
  IAB  \cite{nguyen2020input}& $57.75$  & $99.29$ & $55.25$ & $65.88$ & $54.49$ & $0.22$  & $45.42$ & $97.93$ & $57.19$ & $1.19$ &$47.63$ & $0.91$ \cr
    SSBA  \cite{li2021invisible}& $56.02$  & $98.16$ & $51.54$ & $85.81$ & $52.84$ & $84.57$  & $44.40$ & $99.85$ & $52.92$ & $2.65$ &$53.43$ & $2.21$ \cr 
    \hline
     Average & $54.31$  & $99.01$ & $50.90$ & $87.04$ & $50.33$ & $46.11$  & $44.38$ & $99.18$ & $50.96$ & $2.12$ &$50.12$ & $2.11$ \cr\hline
    
    \end{tabular}  
    
    \caption{Defense methods against common attacks on PreAct-ResNet18 using Tiny-ImageNet.}
    \label{tab:performance_tiny-PreAct-ResNet}  
\end{table*}

\subsection{Backdoor Removal with Shapley Value}
\label{Backdoor removal}
Typically, only a small portion of the total neurons in a model act as backdoor neurons, which are activated only when poisoned samples are provided.
After pruning these poisoned neurons, the backdoor behavior is removed \cite{liu2018fine}. 
Neurons' activation value is the most commonly used method in locating poisoned neurons and the defenders can prune neurons to remove the backdoor according to their activation value \cite{liu2018fine,wang2019neural}.
However, the neuron's activation value is not accurate and some of the normal neurons also have a larger activation value and it will get worse when the number of samples is small.
Thus, it is vital to locate poisoned neurons in a more accurate way.
In this subsection, we leverage Shapley value to locate poisoned neurons and remove the backdoor according to it.
\paragraph{$\triangleright$ Shapley Value.}
In deep neural networks, there are thousands of neurons and complex interactions between them, making it challenging to quantify their contributions to the model's output. 
Shapley value, as an important concept in game theory, can allocate each player's contribution to the outcome \cite{ghorbani2020neuron}.
Shapley value is calculated by the average of each player's marginal value, and Shapley value of player $i$ is expressed as \cite{castro2009polynomial}:
\begin{equation}
   \phi_{i} = \frac{1}{n} \sum\limits_{S\subset N\backslash i}P_{S}\cdot(V(S\cup i)-V(S))
   \label{eq:Shapley}
\end{equation}
where $V$ represents the performance metric, $N=[1,2 \cdots n]$ represents all $n$ players, $S$ represents a subset of $N$ having $s$ players, and $P_{S}=\frac{(n-s-1)! s!}{(n-1)!}$ represents $S$'s relative importance.
In deep neural networks, each player in Equation \ref{eq:Shapley} can be seen as a neuron in the model, and the performance metric $V$ can be used to represent accuracy (ACC) or attack success rate (ASR) in backdoor defense tasks.
Please note that during test time, the partially poisoned samples are unlabeled. 
To address this issue, our method employs the predicted label of the original model as the label for these samples.

\paragraph{$\triangleright$ Estimating Shapley Value.}
Because deep neural networks have thousands of neurons,  directly calculating Shapley value is time-consuming.
To accelerate Shapley calculation, two Shapley estimation acceleration methods are proposed \cite{ghorbani2020neuron,guan2022few}.
First is Monte-Carlo estimation.
From Equation \ref{eq:Shapley}, Shapley value can also be expressed as the average of the marginal value of neurons in all possible orders, expressed as \cite{castro2009polynomial}:
\begin{equation}
   \phi_{i} = E_{\pi \in \Pi} ( V(S_{\pi}^i \cup i)- V(S_{\pi}^i))
   \label{eq:Shapley estimation}
\end{equation}
where $\pi$ represents a random permutation of neurons, $\Pi$ represents permutation set of all neurons, and $S_{\pi}^i$ represents neurons after neuron $i$ in permutation.
With Equation \ref{eq:Shapley estimation}, we use Monte-Carlo estimation to estimate Shapley value.
Furthermore, when pruning neurons during Shapley estimation, models' ACC and ASR  will decrease to near zero after pruning only a small number of neurons, and the marginal values after that are negligible.
Thus we can early stop this pruning process to promote estimation efficiency.
In our experiments, we set a threshold for model pruning, and if the models' performance decreases under this threshold, the pruning is early stopped and another new permutation pruning begins.
In our experiments, we estimate neurons' Shapley value with only 40 average iterations, which is accurate enough for locating the poisoned neurons.

\paragraph{$\triangleright$ Backdoor Removal.}
The Shapley value estimated above indicates the relative importance of neurons to the performance of the model and can be used by defenders to guide neuron pruning. 
Intuitively, facing the partially poisoned sample situation, the defenders leverage the detected poisoned samples and the labels that the original model predicts to estimate the model's ASR Shapley value.   
While the Shapley value is more accurate than model activation, directly pruning the top ASR Shapley value neurons can cause a sharp decrease in accuracy. 
This is due to the fact that the detected poisoned samples may contain normal samples, and the number of poisoned samples is often small.
To address this issue, we estimate the Shapley value of ACC using the entire batch of samples and select neurons with both the top-k ASR Shapley value and the bottom-m ACC Shapley value. Additionally, to prevent neurons from having a small Shapley value due to having both large positive and negative marginal values, we improve Equation \ref{eq:Shapley estimation} and use the mean of the absolute values of the marginal values as the absolute Shapley value, which is expressed as follows:
\begin{equation}
 \hat{\phi_{i}} = E_{\pi \in \Pi} | V(S_{\pi}^i \cup i)- V(S_{\pi}^i)|
\end{equation}
where $\hat{\phi_{i}}$ represents the absolute Shapley value, and $|\cdot |$ represents the absolute value.
Pruning neurons with top-k ASR Shapley value and bottom-m ACC absolute Shapley value, our proposed method locates the poisoned neurons accurately and removes the backdoor with only a small accuracy decrease, only leveraging a batch of partially poisoned data, without the need of fine-tuning the model with clean samples.

\section{Experiment}

\label{different_poisoning_rate}

\begin{table*}[h] \footnotesize
  \centering  
 
    \begin{tabular}{l c c c c c c c c c c c c c c }  
    \hline  
    {Attack}&  
    \multicolumn{2}{c}{Before}&\multicolumn{2}{c}{FP \cite{liu2018fine}} &\multicolumn{2}{c}{DBD \cite{huang2022backdoor}}  &\multicolumn{2}{c}{TTBD-$5\%$} &\multicolumn{2}{c}{TTBD-$10\%$}  &\multicolumn{2}{c}{TTBD-$20\%$}
   
    \cr\cline{2-13}  
    ($\%$)&ACC&ASR&ACC$\uparrow$&ASR$\downarrow$&ACC$\uparrow$&ASR$\downarrow$&ACC$\uparrow$&ASR$\downarrow$&ACC$\uparrow$&ASR$\downarrow$&ACC$\uparrow$&ASR$\downarrow$\cr 
    \hline  
    BadNet  \cite{gu2017badnets} & $90.53$  & $84.45$ & $90.35$ & $79.58$ & $57.54$ & $2.48$ & $84.67$ & $3.68$ & $87.11$ & $1.63$ &$88.30$ & $1.60$ \cr
    Blended \cite{chen2017targeted}  & $90.81$  & $88.31$ & $90.62$ & $87.90$& $55.24$ & $6.21$ & $82.34$ & $5.96$ & $86.97$ & $4.81$ &$85.32$ & $1.94$ \cr
    SIG \cite{barni2019new} & $90.75$  & $82.93$ & $90.60$ & $83.46$ & $55.92$ & $0.00$ & $88.61$ &$0.99$ & $89.48$ & $2.64$ &$89.23$ & $0.98$ \cr
    LF \cite{zeng2021rethinking}  & $88.88$  & $93.43$ & $88.79$ & $92.39$& $56.83$ & $10.21$ & $88.02$ & $2.28$ & $87.04$ & $1.68$ &$88.18$ & $1.92$ \cr
    WaNet \cite{nguyen2021wanet}  & $89.29$  & $85.63$ & $89.67$ & $72.07$ & $55.80$ & $15.22$ & $88.35$ & $3.54$ & $88.58$ & $1.82$ &$88.47$ & $1.46$ \cr 
     IAB \cite{nguyen2020input}   & $89.24$  & $71.79$ & $89.73$ & $77.43$ & $57.21$ & $13.10$  & $87.93$ & $1.28$ &$84.77$ & $3.92$ &$86.22$ & $1.67$ \cr
    SSBA \cite{li2021invisible}  & $89.66$  & $91.28$ & $89.59$ & $88.97$ & $58.45$ & $12.20$ & $89.14$ & $1.80$ &$89.02$ & $1.98$ &$88.04$ & $2.58$ \cr

    \hline
    Average & $89.88$  & $85.40$ & $89.91$ & $83.11$ & $56.71$ & $8.49$ &  $87.00$ & $2.79$ &  $87.57$ & $2.64$ &$87.68$ & $1.74$ \cr\hline
    
    \end{tabular}  
    \caption{Performance of TTBD-DDP with different poisoning rates.}
    \label{tab:performance_posion_rate}  
\end{table*} 

\begin{table*}[h] \footnotesize
  \centering  
  
    \begin{tabular}{c c c c c c c c c c c c c}  
    \hline  
    {Attack}&  
    \multicolumn{2}{c}{Before}&\multicolumn{2}{c}{FP \cite{liu2018fine}} & \multicolumn{2}{c}{DBD \cite{huang2022backdoor}} &\multicolumn{2}{c}{TTBD-50} &\multicolumn{2}{c}{TTBD-100}  &\multicolumn{2}{c}{TTBD-200}
   
    \cr\cline{2-13}  
    ($\%$)&ACC&ASR&ACC$\uparrow$&ASR$\downarrow$&ACC$\uparrow$&ASR$\downarrow$&ACC$\uparrow$&ASR$\downarrow$&ACC$\uparrow$&ASR$\downarrow$&ACC$\uparrow$&ASR$\downarrow$\cr 
    \hline  
    BadNet \cite{gu2017badnets}& $90.53$  & $84.45$ & $90.35$ & $79.58$ & $57.54$ & $2.48$& $85.30$ & $1.24$ & $87.11$ & $1.63$ &$88.42$ & $1.26$ \cr
    Blended  \cite{chen2017targeted} & $90.81$  & $88.31$ & $90.62$ & $87.90$ & $55.24$ & $6.21$& $86.07$ & $4.40$ & $86.97$ & $4.81$ &$87.09$ & $2.95$ \cr
    SIG \cite{barni2019new}  & $90.75$  & $82.93$ & $90.60$ & $83.46$ & $55.92$ & $0.00$& $88.68$ & $1.87$ & $89.48$ & $2.64$ &$89.50$ & $1.00$ \cr
    LF \cite{zeng2021rethinking} & $88.88$  & $93.43$ & $88.79$ & $92.39$ & $56.83$ & $10.21$& $87.42$ & $2.21$ & $87.04$ & $1.68$ &$87.73$ & $2.82$ \cr
    WaNet  \cite{nguyen2021wanet}& $89.29$  & $85.63$ & $89.67$ & $72.07$& $55.80$ & $15.22$ & $86.77$ & $5.44$ & $88.58$ & $1.82$ &$87.69$ & $2.53$ \cr
    IAB \cite{nguyen2020input} & $89.24$  & $71.79$ & $89.73$ & $77.43$ & $57.21$ & $13.10$  & $86.10$ & $1.46$ & $84.77$ & $3.92$ &$86.02$ & $1.47$ \cr
    SSBA  \cite{li2021invisible}& $89.66$  & $91.28$ & $89.59$ & $88.97$ & $58.45$ & $12.20$ & $87.82$ & $1.71$ & $89.02$ & $1.98$ &$89.14$ & $2.36$ \cr \hline
    Average &  $89.88$  & $85.40$ & $89.91$ & $83.11$ & $56.71$ & $8.49$& $86.88$  & $2.62$ & $87.57$ & $2.64$ & $87.94$& $2.06$  \cr
    \hline
    
    \end{tabular}  
    
    \caption{Performance of TTBD-DDP with different batch sizes.}
    \label{tab:performance_batchsize} 
\end{table*} 

\begin{table*}[h] \footnotesize
  \centering  
  
    \begin{tabular}{l c c c c c c c c c c c c}  
    \hline  
    \multicolumn{2}{c}{Attack($\%$)}&  
    BadNet \cite{gu2017badnets}&Blended \cite{chen2017targeted}& SIG \cite{barni2019new}&LF \cite{zeng2021rethinking}&WaNet \cite{nguyen2021wanet} &IAB \cite{nguyen2020input} & SSBA  \cite{li2021invisible}& Average
    \cr\hline

    \multirow{2}{*}{Before} & ACC & 91.23 & 93.76 & 91.45 & 93.76 & 91.48 & 91.09 & 93.43 & 92.31 \cr
  & ASR & 90.22 & 94.88 & 91.47 & 86.74 & 89.91 & 90.61 &73.44 & 88.18 \cr\hline
    \multirow{2}{*}{TTBD-SPARSE} & ACC  & 83.05 & 84.88 & 88.27 & 83.69 & 87.33 & 82.69 & 81.62 & 84.50  \cr
     & ASR& 9.67 & 8.98 & 1.50 & 7.60 & 0.89 & 6.97 & 4.28& 5.70   \cr
   
    \hline
    
    \end{tabular}  
    
    \caption{Performance of TTBD under extremely low poisoning rate.}
    \vspace{-3mm}
    \label{tab:performance_sparse} 
\end{table*}

\subsection{Setup}
\paragraph{$\triangleright$ Datasets and Model Architectures.}
We evaluate various backdoor defense methods on three common datasets used in backdoor defense, CIFAR10, CIFAR100 \cite{krizhevsky2009learning}, and Tiny-ImageNet \cite{le2015tiny}.
Furthermore, following the BackdoorBench \cite{wu2022backdoorbench}, we also evaluate all backdoor defense methods on three common model architectures, PreAct-ResNet18 \cite{he2016identity}, VGG19 \cite{simonyan2014very}, and DenseNet161 \cite{huang2017densely}.

\paragraph{$\triangleright$ Attack Settings.}
We consider 7 popular state-of-the-art backdoor attacks: BadNets \cite{gu2017badnets}, Blended \cite{chen2017targeted}, SIG \cite{barni2019new}, LF \cite{zeng2021rethinking}, WaNet \cite{nguyen2021wanet}, IAB \cite{nguyen2020input},and SSBA \cite{li2021invisible}.
We follow the default configuration in BackdoorBench \cite{wu2022backdoorbench} for a fair comparison.
As for the poisoning rate, to test different defense methods' performance on different poisoning rates, we set the poisoning rate to $1\%$ for CIFAR10 and $10\%$ for Tiny-ImageNet. 
To ensure the integrity of our experiments, we set the poisoning rate to $10\%$ for WaNet on CIFAR10 using both PreAct-ResNet and VGG, and for LF and SSBA on CIFAR10 using VGG, as these attacks are not able to successfully inject a backdoor at $1\%$ poisoning rate ($1\%$ poisoning rate injection causes a very low ASR). 
Additionally, since SIG cannot perform an attack on Tiny-ImageNet with poisoning rates of either $1\%$ or $10\%$, we did not use SIG on Tiny-ImageNet.

\paragraph{$\triangleright$ Defense Settings.}
Since there have been no prior test-time backdoor defense methods based on model repairing, we compare our method with previous state-of-the-art model-repairing-based backdoor defense methods, including Fine Pruning (FP) \cite{liu2018fine} and Adversarial Neuron Pruning (ANP) \cite{wu2021adversarial}. 
Additionally, to fully verify the effectiveness of our method, we also include a state-of-the-art training-stage backdoor defense, DBD \cite{huang2022backdoor}.
We follow the default configuration in BackdoorBench \cite{wu2022backdoorbench} for a fair comparison.
To ensure the effectiveness of the compared backdoor defense methods, following BackdoorBench, we have specified that FP and ANP have access to $5\%$ of the benign training data, and DBD can access the full poisoned training dataset and control model's training process.
Additionally, we also compare with ShapPruning \cite{guan2022few} in Appendix.
We set the partially poisoned data's poisoning rate to $10\%$ in our test-time backdoor defense setting, and the image batch used in TTBD only consists of 100 partially poisoned samples. 
Furthermore, we study the influence of the poisoning rates and the batch sizes on our TTBD's performance in the influence of poisoning rates and batch sizes subsection. 
Additionally, for poisoned sample detection, we have selected the top 10 detected samples in TeCo and the top 6 detected samples in DDP as our identified poisoned samples. 
We have used these samples to estimate the ASR Shapley value of neurons and remove the backdoor in the models.

\paragraph{$\triangleright$ Evaluation Metric.}
We use two commonly used metrics to measure the effectiveness of different backdoor defense methods: accuracy on clean samples (ACC) and attack success rate on poisoned samples (ASR).
The ultimate goal of defenders is to leverage various methods to achieve a no-backdoor model with high ACC and low ASR.

\subsection{Experiment Results}

Tables \ref{tab:performance_CIFAR10-PreAct-ResNet}, \ref{tab:performance_CIFAR10-VGG}, and \ref{tab:performance_tiny-PreAct-ResNet} demonstrate the performance of TTBD-TeCo and TTBD-DDP across different model architectures (PreAct-ResNet18 and VGG19) and datasets (CIFAR10 and Tiny-ImageNet), as evaluated against 7 different state-of-the backdoor attacks.
Additionally, we have also conducted experiments on DenseNet161 and CIFAR100 in Appendix.
In the tabeles, Before represents the original backdoor model without model defense mechanisms, while FP, ANP, DBD, TTBD-TeCo, and TTBD-DDP represent the repaired model with the specific backdoor defense. 
In most cases, our TTBD backdoor defense methods remove backdoor in models with a small decline in accuracy (around $2\%$ average accuracy decline on CIFAR10). 
In Table \ref{tab:performance_CIFAR10-VGG}, because ANP can not work on VGG (VGG19 model does not have batch normalization layers), we use '-' to fill in the blank.
Due to the possibility of detection errors in some samples during TeCo and DDP, a small percentage of poisoned samples may be mistakenly identified as clean samples. 
This makes it difficult for the defender to fine-tune the model using the detected clean samples directly, as the presence of even a very small percentage of poisoned samples can lead to the backdoor being retained in the model.
Typically, most previous model repairing backdoor defense methods need clean samples to fine-tune the backdoor model and maintain the models' accuracy.
Although without clean samples to fine-tune the model, our TTBD-TeCo and TTBD-DDP can effectively remove the backdoor by pruning a small percentage of neurons, and our backdoor removal procedure removes the backdoor (using top-k ASR Shapley values) with a small decline in accuracy (using bottom-m ACC absolute Shapley values).  
In Table \ref{tab:performance_tiny-PreAct-ResNet}, there is a slightly greater decrease in accuracy on Tiny-ImageNet by our TTBD-based methods compared with the results on CIFAR10.
This is due to the lack of fine-tuning with clean data, as models trained on Tiny-ImageNet have fewer redundant neurons than models trained on CIFAR10. 
And if the defender can obtain access to some clean samples for fine-tuning, the accuracy of the models can be recovered.
While TTBD-TeCo and TTBD-DDP yield comparable backdoor defense results on PreAct-ResNet using both CIFAR10 and Tiny-ImageNet, TTBD-TeCo is not effective on VGG models. This is because TeCo's detection is sensitive to variations in model architectures, resulting in an AUC of approximately 0.5 on VGG models.
In contrast, TTBD-DDP successfully removes backdoor attacks across various model architectures and datasets with pruning a very small portion of neurons, which is shown in Appendix.

\begin{figure*}[!htb]
		\centering
		\small
		\setlength\tabcolsep{1mm}
		\renewcommand\arraystretch{0.1}
		\begin{tabular}{ccccc}
			\includegraphics[width=0.18\linewidth]{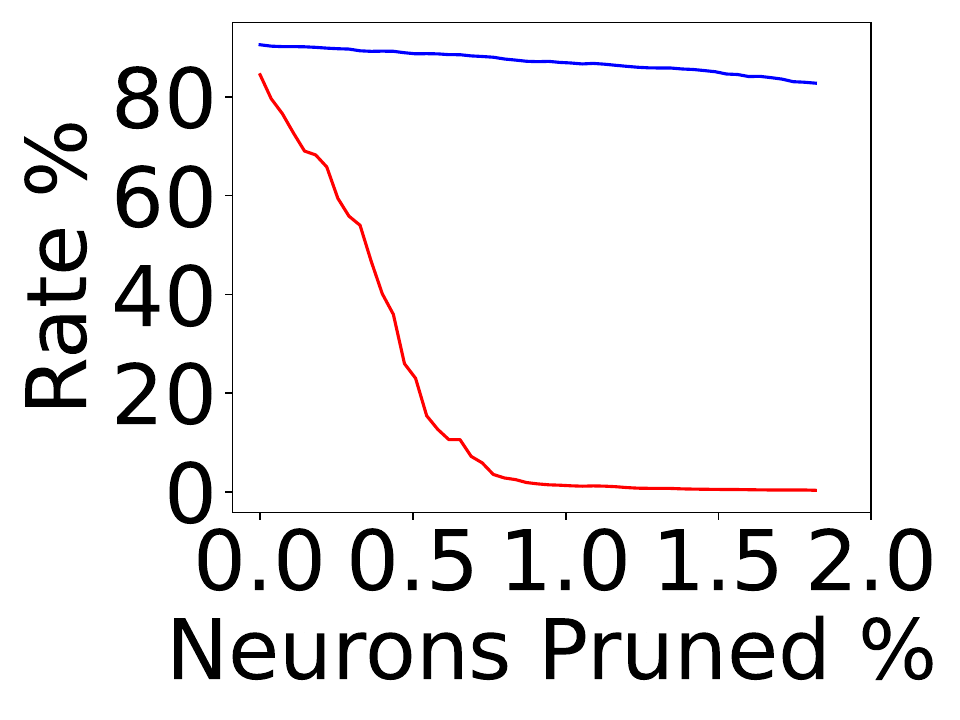} &
			\includegraphics[width=0.18\linewidth, clip]{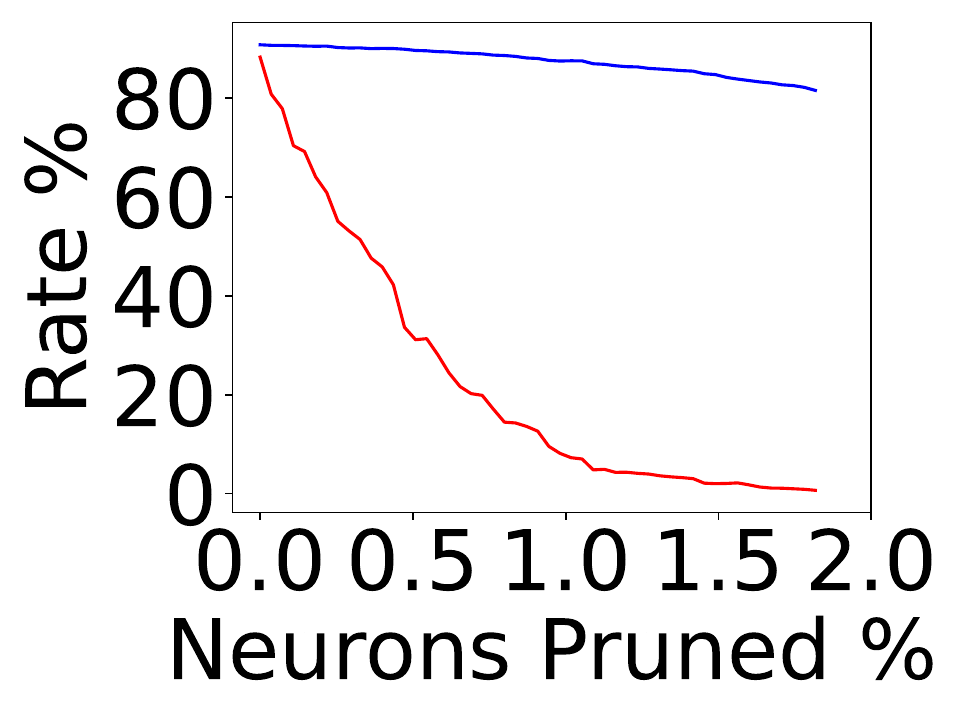} & 
                 \includegraphics[width=0.18\linewidth, clip]{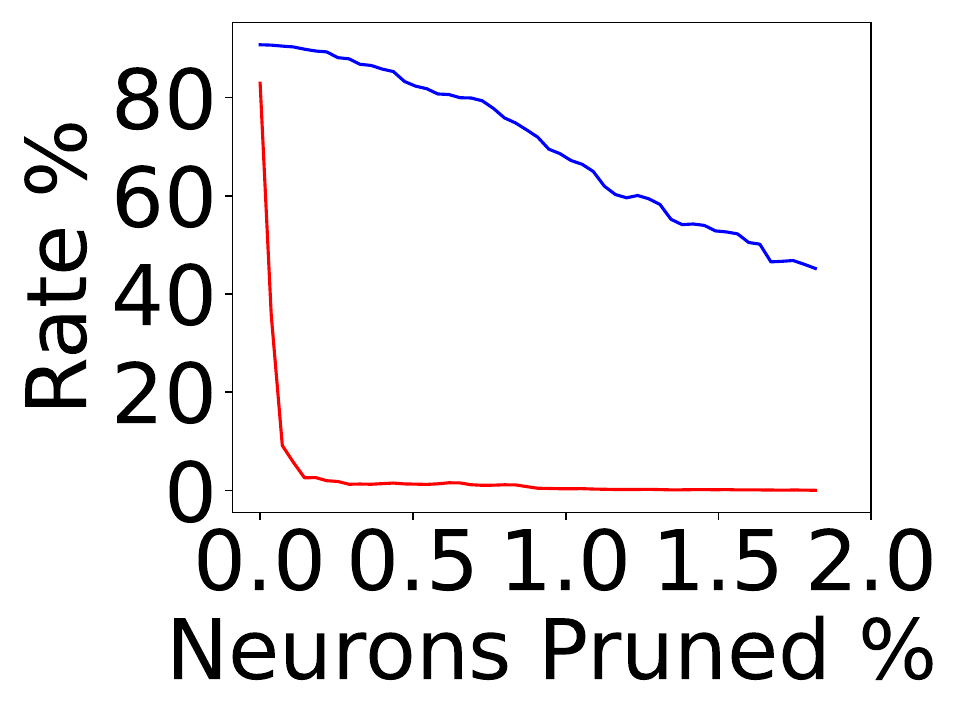} & 
                  \includegraphics[width=0.18\linewidth, clip]{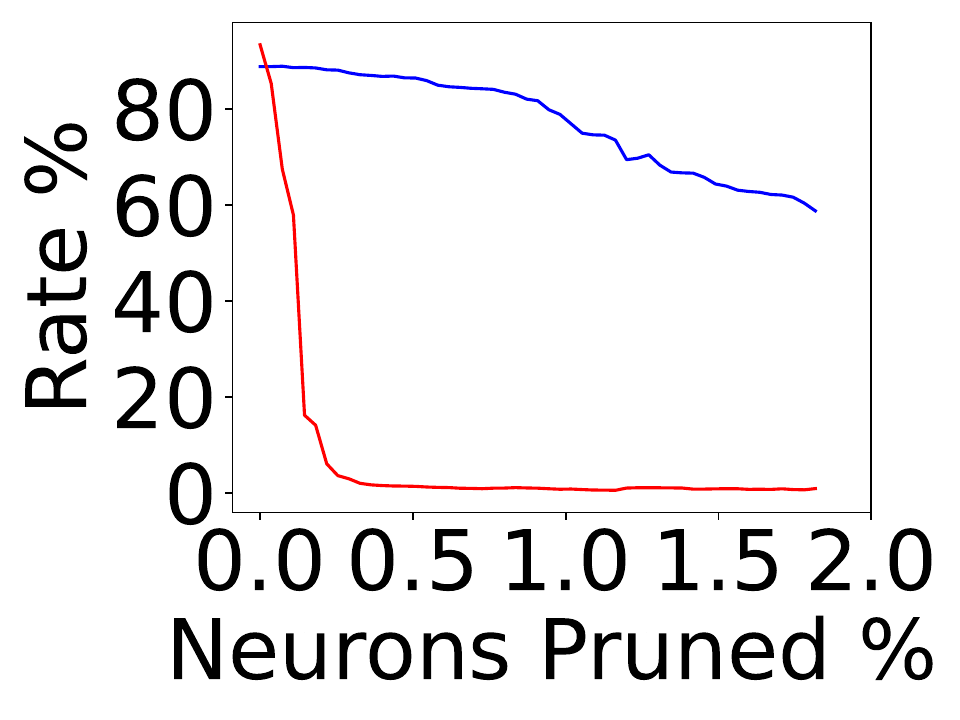} & 
			\includegraphics[width=0.18\linewidth, clip]{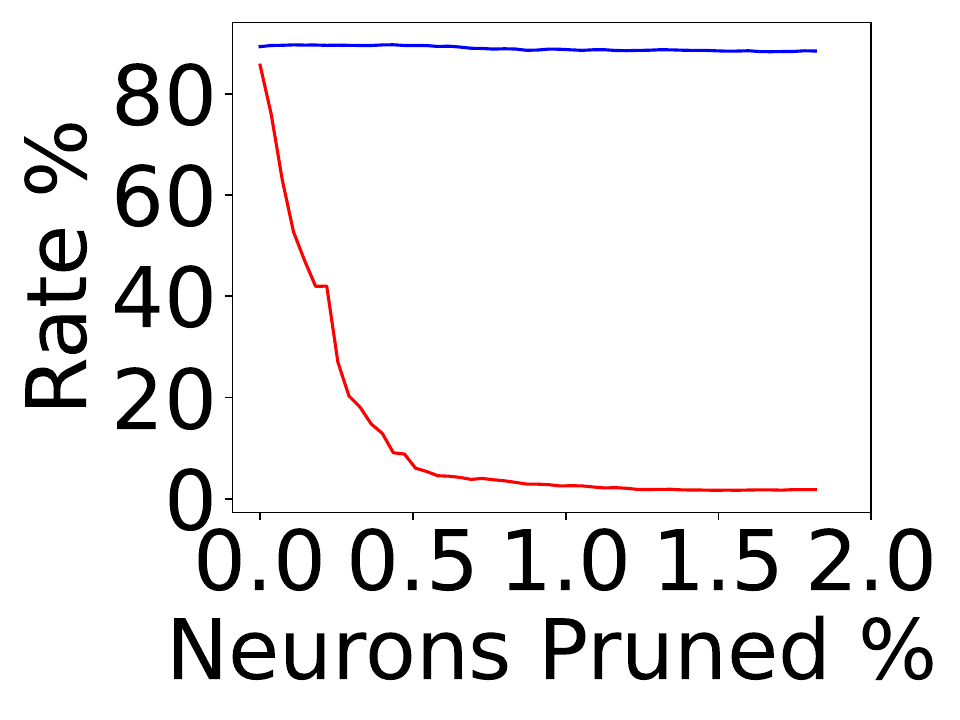} \\
			\\[0.5mm]
			(a) DDP-BadNet & (b) DDP-Blended  & (c) DDP-SIG & (d) DDP-LF & (e) DDP-WaNet
		\end{tabular}
  \begin{tabular}{ccccc}
			\includegraphics[width=0.18\linewidth]{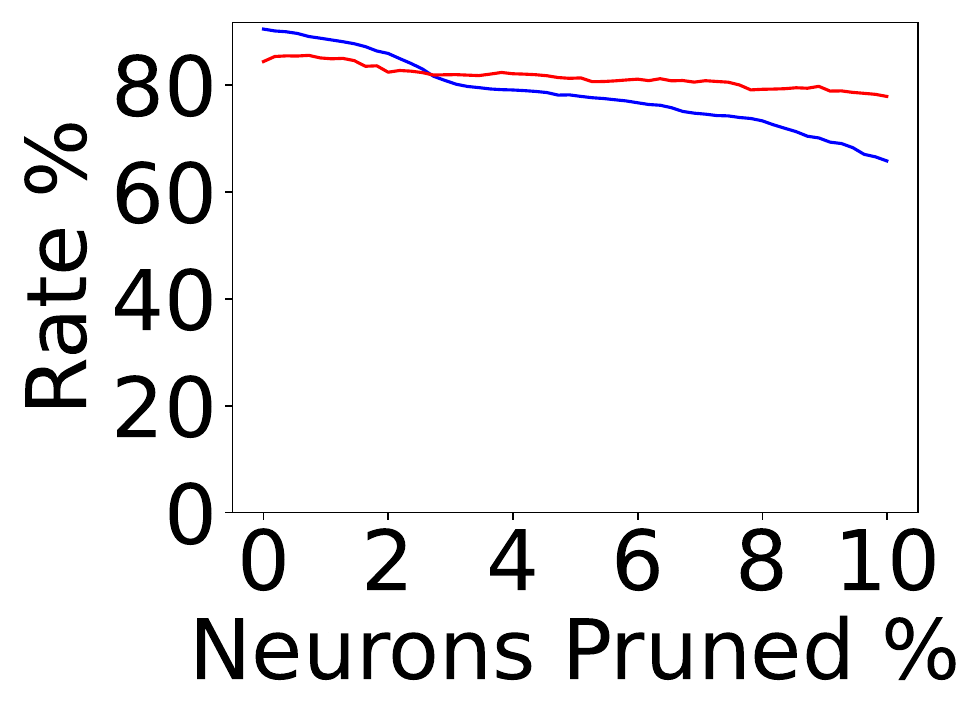} &
			\includegraphics[width=0.18\linewidth, clip]{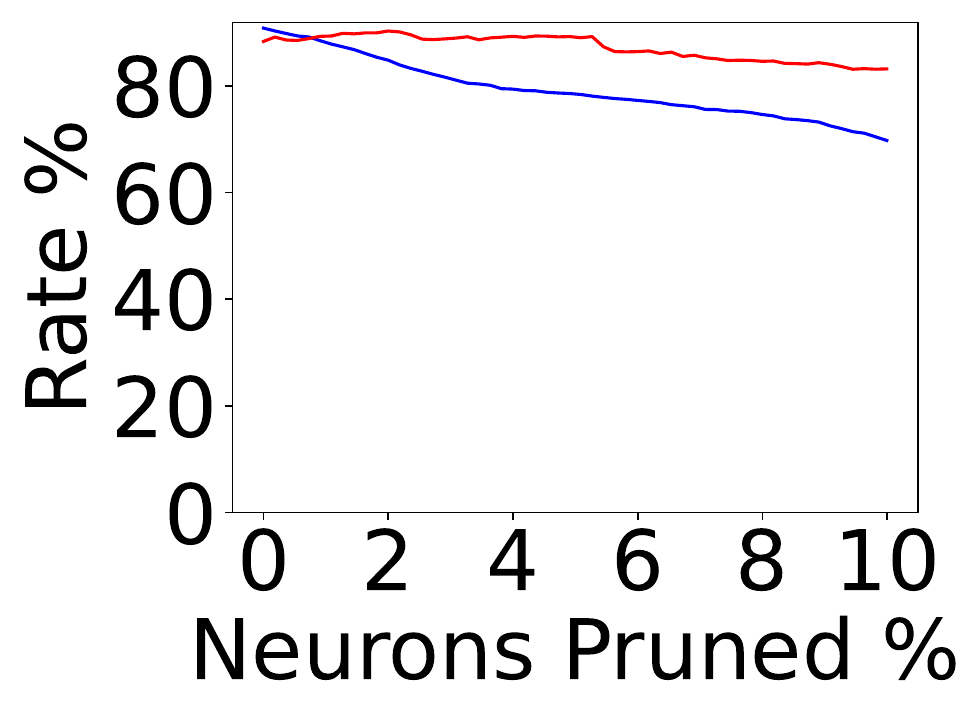} & 
                 \includegraphics[width=0.18\linewidth, clip]{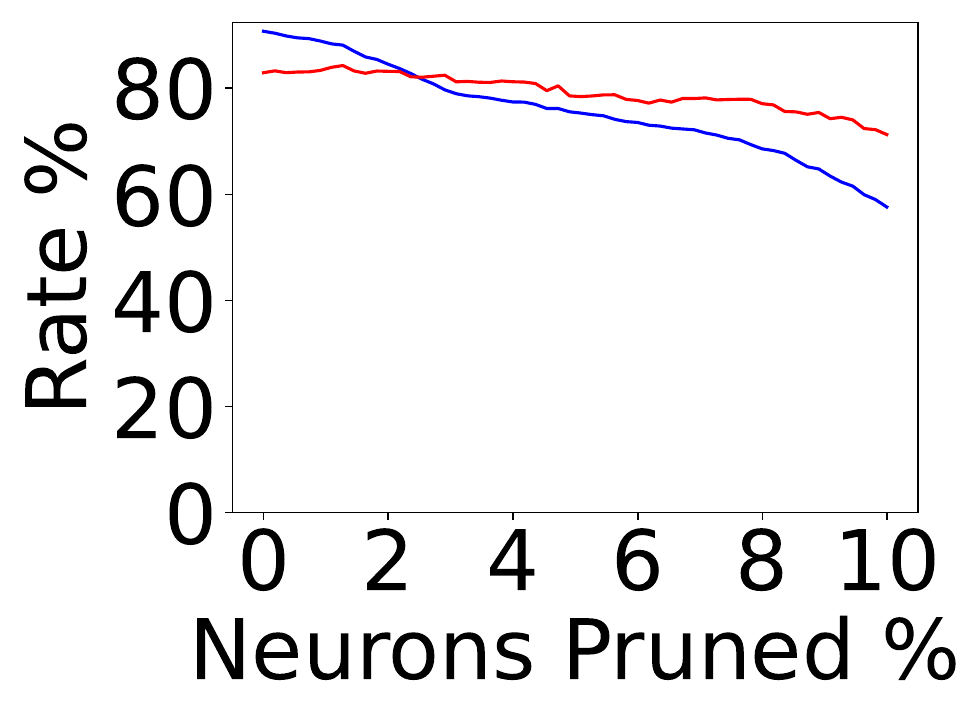} & 
                  \includegraphics[width=0.18\linewidth, clip]{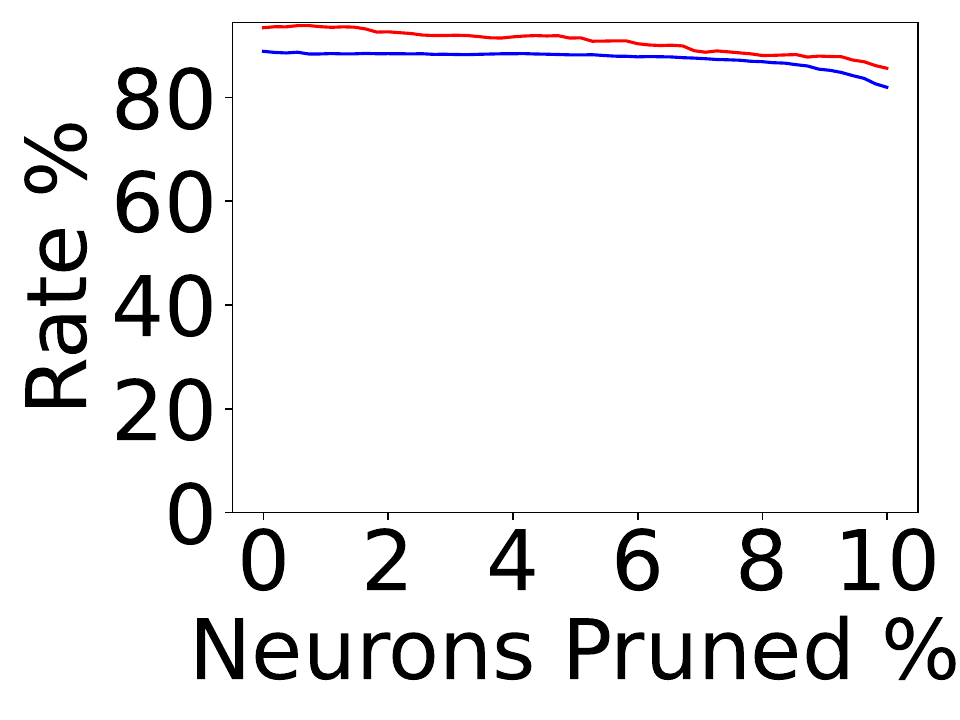} & 
			\includegraphics[width=0.18\linewidth, clip]{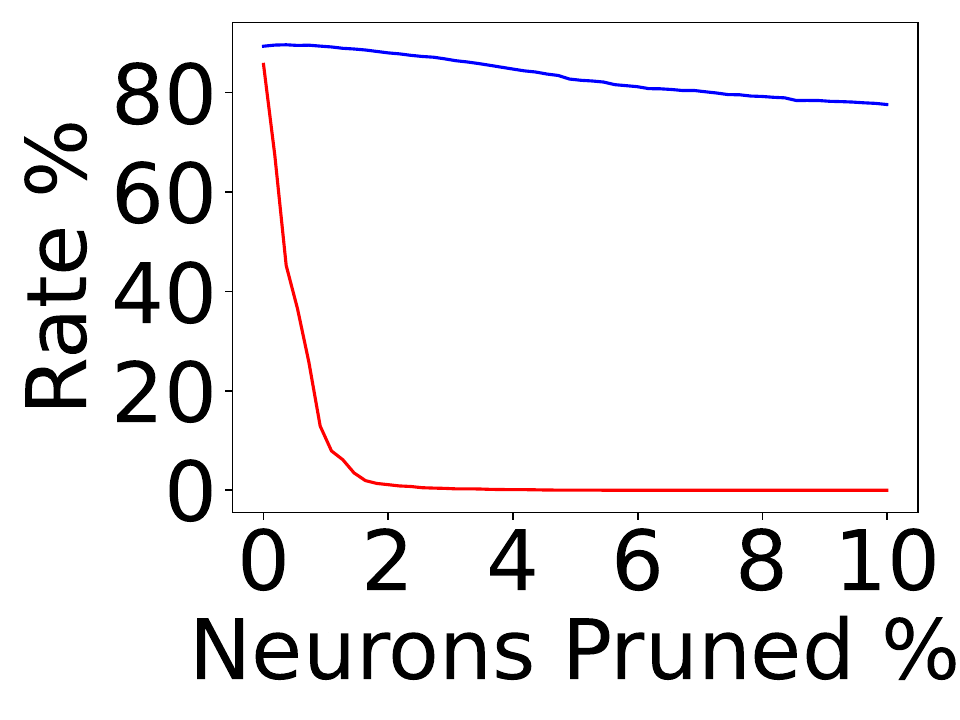} \\
			\\[0.5mm]
			(f) TeCo-BadNet & (g) TeCo-Blended  & (h) TeCo-SIG & (i) TeCo-LF & (j) TeCo-WaNet
		\end{tabular}
        \vspace{1pt}
		\caption{TTBD's ACC and ASR fluctuation during neuron pruning on VGG19. Red lines represent ASR and blue lines represent ACC.}
		\label{fig:auc_sample_num}
\end{figure*}

On the other hand, the compared methods have their weaknesses and are not as successful in removing backdoor attacks from models.
For instance, FP \cite{liu2018fine} fails because neuron activation is not an accurate way to identify the poisoned neurons. As a result, pruning neurons based on the smallest activation is unable to pinpoint and eliminate the poisoned neurons responsible for the backdoor attack.
Additionally, ANP \cite{wu2021adversarial} leverages neurons which are sensitive to adversarial training and masks them, but this approach leads to significant accuracy reductions in the models. 
DBD \cite{huang2022backdoor} leverages self-supervised learning to train backbone models, train the fully connected layers, and removes labels of low-credit samples to fine-tune the source models.
Table \ref{tab:performance_CIFAR10-PreAct-ResNet} demonstrates that while DBD is able to remove backdoors from the models, the accuracy of the resulting models is significantly lower than the accuracy of the original backdoor models.
In comparison, TTBD-DDP reduces the backdoor to an average ASR of $2.32\%$ with only a $2.64\%$ accuracy decrease, and TTBD-TeCo reduces the backdoor to an average ASR of $1.55\%$ with only a $2.96\%$ accuracy decrease on PreAct-ResNet18 using CIFAR10. 

\begin{table*}[h] \footnotesize
  \centering  
 
    \begin{tabular}{l c c c c c c c c c c}  
    \hline  
    {Attack}&  
    \multicolumn{2}{c}{Before}&\multicolumn{2}{c}{TTBD-RAND} &\multicolumn{2}{c}{TTBD-ACT} &\multicolumn{2}{c}{TTBD-DDP}
   
    \cr\cline{2-9}  
    ($\%$)&ACC&ASR&ACC$\uparrow$&ASR$\downarrow$&ACC$\uparrow$&ASR$\downarrow$&ACC$\uparrow$&ASR$\downarrow$\cr 
    \hline  
    BadNet \cite{gu2017badnets} & $91.23$  & $90.22$ & $82.30$ & $10.90$ & $88.58$ & $51.53$ & $88.50$ & $2.51$ \cr
    Blended  \cite{chen2017targeted}& $93.76$  & $94.88$ & $85.29$ & $92.90$ & $88.65$ & $3.28$ & $88.53$ & $2.24$ \cr
    SIG \cite{barni2019new}  & $91.45$  & $91.47$ & $82.92$ & $81.08$ & $83.30$ & $18.10$ & $89.59$ & $2.77$ \cr
    LF  \cite{zeng2021rethinking} & $93.76$  & $86.74$ & $85.56$ & $91.88$ & $85.32$ & $35.74$ & $90.47$ & $2.72$ \cr
    WaNet \cite{nguyen2021wanet} & $91.48$  & $89.91$ & $83.21$ & $78.78$ & $91.02$ & $0.50$ & $91.07$ & $0.78$ \cr
     IAB \cite{nguyen2020input}  & $91.09$  & $90.61$ & $82.09$ & $92.63$ & $89.95$ & $1.47$ & $88.47$ & $1.97$ \cr
    SSBA \cite{li2021invisible}  & $93.43$  & $73.44$ & $84.71$ & $8.42$ & $90.35$ & $3.97$ & $91.09$ & $3.28$ \cr
    \hline
    Average & $92.31$  & $88.18$ & $83.73$ & $65.23$ & $88.17$ & $16.37$ & $89.67$ & $2.32$ \cr\hline
    \end{tabular}  
    
    \caption{Ablation study against five common attacks on PreAct-ResNet18 using CIFAR10.}
    \label{tab:ablation-study}  
\end{table*}

\subsection{Sensitivity Analysis}
In this subsection, we analyze the influence of different poisoning rates and batch sizes on TTBD-DDP on VGG19 using CIFAR10. 
Table \ref{tab:performance_posion_rate} demonstrates TTBD-DDP's performance across different poisoning rates of the partially poisoned image batch.
Specifically, TTBD-$5\%$, TTBD-$10\%$, and TTBD-$20\%$ represent TTBD-DDP's performance using the image batch with $5\%$, $10\%$, and $20\%$ poisoning rates, respectively, with the batch size of 100.
With the increase of the poisoning rates, TTBD-DDP's performance rises and TTBD-DDP removes different types of backdoor attacks with a low poisoning rate.
Table \ref{tab:performance_batchsize} demonstrates TTBD-DDP's performance across different batch sizes of the image batch, where TTBD-50, TTBD-100, and TTBD-200 represent TTBD-DDP's performance using $50$, $100$, $200$ partially poisoned samples with the poisoning rate of $10\%$.
With the increase of the batch sizes, TTBD-DDP's performance rises and TTBD-DDP removes all the seven backdoor attacks with only 50 partially poisoned samples.
We also consider a sparse condition for the backdoor attackers, where they poison at a very low poisoning rate, and only some of the batches are poisoned. 
Following TTBD-SPARSE in \ref{TTBD-SPARSE}, we leverage DDP and TeCo as a dual detection method to detect and remove backdoor, and the results are demonstrated in Table \ref{tab:performance_sparse}.
TTBD-SPARSE represents the defenders removing the backdoor by aggregating 20 batches of images and leveraging TeCo and DDP as a dual detection approach when the attackers attack with a very low poisoning rate $1\%$ and sparsely (some of the batches are totally clean).

\subsection{Ablation Study}

In this subsection, we will take poisoned sample detection and backdoor neuron locating into consideration to validate our method's effectiveness.
In Table \ref{tab:ablation-study}, TTBD-RAND represents randomly selecting samples from the partially poisoned samples and using Shapley estimation to prune the poisoned neurons, and TTBD-ACT represents leveraging samples detected by DDP and using neurons' activation to locate poisoned neurons.
Randomly chosen samples are not relevant to the backdoor behavior, and thus, the models after TTBD-RAND have an average $22.95\%$ ASR decrease but also have an average $8.58\%$ ACC decrease.
Additionally, due to the fact that neurons' activation on poisoned samples does not directly correspond to their importance in terms of ASR, the TTBD-ACT approach exhibits a lower ACC-ASR compared to TTBD-TeCo and TTBD-DDP when facing different types of attacks.
Moreover, the presence of clean samples mixed in with the detected poisoned samples and the limited number of detected samples present challenges for TTBD-ACT.
As a result of these factors, TTBD-ACT is unable to completely remove the backdoor effect in the models, and the average ASR remains at $16.37\%$.

\section{Conclusion and Limitation}

This paper introduces a test-time backdoor defense framework called TTBD, which leverages a two-stage framework to detect and remove the backdoor from the poisoned models with only a batch of partially poisoned samples.
To detect the poisoned samples from the partially poisoned data, a backdoor sample detection method DDP is proposed. 
DDP leverages the prediction changes during pruning to accurately detect poisoned samples.
After poisoned sample detection, we leverage Shapley estimation to prune the backdoor-related neurons.
Using the TTBD framework, two methods, TTBD-DDP and TTBD-TeCo, successfully remove seven state-of-the-art backdoor attacks using only a batch of partially poisoned data across different model architectures and datasets.
Furthermore, TTBD demonstrates the ability to remove backdoor in models with varying batch sizes or poisoning rates.
Considering that the performance of TTBD is influenced by the accuracy of poisoned sample detection, and DDP can be further improved, our future research will focus on developing more precise methods for detecting poisoned samples.


{
    \small
    \bibliographystyle{ieeenat_fullname}
    \bibliography{main}
}


\end{document}